# A Vision Architecture

Christoph von der Malsburg
*Frankfurt Institute for Advanced Studies*
malsburg@fias.uni-frankfurt.de

Abstract:    We are offering a particular interpretation (well within the range of experimentally and theoretically accepted notions) of neural connectivity and dynamics and discuss it as the data-and-process architecture of the visual system. In this interpretation the permanent connectivity of cortex is an overlay of well-structured networks, "nets", which are formed on the slow time-scale of learning by self-interaction of the network under the influence of sensory input, and which are selectively activated on the fast perceptual time-scale. Nets serve as an explicit, hierarchically structured representation of visual structure in the various sub-modalities, as constraint networks favouring mutually consistent sets of latent variables and as projection mappings to deal with invariance.

## INTRODUCTION

The performance of human visual perception is far superior to that of any computer vision system and we evidently still have much to learn from biology. Paradoxically, however, the functionality of neural vision models is worse, much worse, than that of computer vision systems. We blame this shortfall on the commonly accepted neural data structure, which is based on the single neuron hypothesis (Barlow, 1972). We propose here a radically new interpretation of neural tissue as data structure, in which the central role is played by structured neural nets, which are formed in a slow process based on synaptic plasticity and which can be activated on the fast psychological time-scale. This data structure and the attendant dynamic processes make it possible to formulate a vision architecture into which, we argue, many of the algorithmic processes developed in decades of computer vision can be adapted.

## STRUCTURED NETS

Although we live in a three-dimensional world, biological vision is, to all we know, based on "2.5-dimensional" representations, that is, two-dimensional views enriched with local depth information. The vision modality has many sub-modalities – texture, colour, depth, surface curvature, motion, segmentation, contours, illumination and more. All of these can naturally be represented in terms of local features tied together into two-dimensional "nets" by active links. Nets are naturally embedded in two-dimensional manifolds and have short-range links between neurons. Neural sheets, especially also the primary visual cortex, can support a very large number of nets by sparse local selection of neurons, which are then linked up in a structured fashion (see Fig. 1). Given the cell-number redundancy in primary visual cortex (exceeding geniculate numbers by estimated factors of 30 or 50) there is much combinatorial space to define many nets. These nets are formed by statistical learning from input and by dynamic self-interaction. In this way a distributed memory for local texture in the various modalities can be stored already in retinal coordinates, that is, in primary visual cortex.

We would like to stress here the contrast of this mode of representation to the current paradigm. To cope with the structure of the visual world, a vision system has to represent a hierarchy of sub-patterns, "features". In standard multi-layer perceptrons (for an early reference see Fukushima, 1980) all features of the hierarchy are represented by neurons. What we are proposing here amounts to replacing units as representatives of complex features by local pieces of net structure tying together low-level feature neurons (or "texture elements", neurons representing the elementary features that are found in neurophysiological experiments in primary visual cortex). This has a number of decisive advantages. First, structured nets represent visual patterns explicitly, as a two-dimensional arrangement of local texture elements. Second, as alluded to above, large numbers of nets can be implemented on a comparatively narrow neural basis in a combinatorial fashion. Third, partial identities of different patterns are taken care of by partial identity of the representing pieces of net structure. Fourth, a whole hierarchy of features can be represented in a flat structure already in primary visual cortex (a shade of neurophysiological evidence for the lateral connections between neurons surfaces in the form of non-classical receptive fields, see Allman et al. 1985). Fifth, nets that are homeomorphic to each other (i.e., can be put into neuron-to-neuron correspondence such that connected neurons in one net correspond to connected neurons in the other) can activate each other directly, without this interaction having to be taught, see below.

## ACTIVATION OF NETS

Once local net structure has been established by learning and self-interaction, the activation by visual input takes the following form (see Fig. 1). The sensory input selects local feature types. Each feature type is (at least in a certain idealization) represented redundantly by a number of neurons with identical receptive fields. Sets of such input-identical redundant neurons form "units". Within a unit there is an inhibitory system inducing winner-take-all (WTA) dynamics (only one or a few of the redundant neurons surviving after a short time). The winners in this process are those neurons that form part of a net, that is, whose activity is supported by lateral, recurrent input.

This process of selection of the input-activated neurons that happen to be laterally connected as a net is an important type of implementation of dynamic links: although the connections are actually static, nets are dynamically activated by selection of net-bearing neurons. For another type see below.

Local pieces of net structure can be connected like a continuous mosaic into a larger net. This may be compared to the image-compression scheme in which the texture within local blocks of an image is identified with a code-book entry (only the identifying number of the code-book entry being transmitted).

## GENERATION OF NETS

Net structure in primary visual cortex is shaped by two influences, input statistics and self-interaction. One may assume that the genetically generated initial structure has random short-range lateral connections. In a first bout of organization receptive fields of neurons are shaped by image statistics, presumably under the influence of a sparsity constraint (Olshausen and Field, 1996). In this period the WTA inhibition may not yet be active, letting neurons in a unit develop the same receptive field. Then, the network becomes sensitive to the statistics of visual input within somewhat larger patches (the scale being set by the range of lateral connections) and pieces of net structure are formed by synaptic plasticity strengthening connections between neurons that are often co-activated and WTA-selected, while net structure is optimized by the interplay between (spontaneous or induced) signal generation and Hebbian modification of synaptic strengths under the influence of a sparsity constraint .

## MODALITIES

Different sub-modalities (texture, colour, depth, motion, ..) form their own systems of net structure, that is, representations of local patterns that are statistically dominant in the sensory input. Each modality is invariant to the others and has its own local feature space structure with its own dimensionality, three for colour, two for in-plane motion, one for (stereo-)depth, perhaps 40 for grey-level texture and so on. Different values of a given feature dimension are represented by different neurons, or rather units containing a number of value-identical neurons. Different value-units of the same feature dimension, forming a "column", inhibit each other, again in WTA fashion.

# LATENT VARIABLES

Several units standing for different values of a sub-modality feature may be simultaneously active to varying degree. They may be seen as representing different hypotheses as to the actual value of the feature dimension. These activities thus represent heuristic uncertainty, which during the perceptual process needs to be reduced to certainty. In distinction to computer graphics, realized as a deterministic process proceeding from definite values of all involved variables determining a scene, vision is an inverse problem, in which these values first have to be found in a heuristic process that is inherently non-deterministic. The initially unknown quantities are called latent variables. The task of the perceptual process is the iterative reduction of the heuristic uncertainty of latent variables ("perceptual collapse"), which is possible by the application of consistency constraints and known memory patterns.

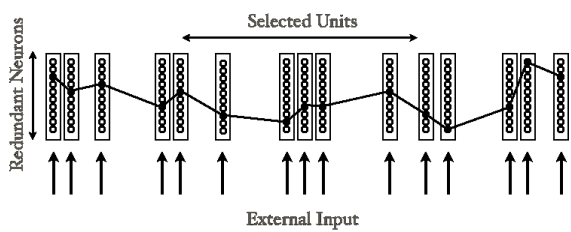

Figure 1: Combinatorially many nets co-exist within a cortical structure (schematic). Units (vertical boxes) are sets of neurons with identical receptive field. Visual input selects a sparse subset of units (vertical arrows). Neurons within units have WTA dynamics. The winner neurons are those that are supported by lateral connections from neurons in other selected units. Lateral connections form a net structure. A neuron can be part of several nets, thus, many nets can co-exist.

# CONSISTENCY CONSTRAINS

Whereas the winner neurons within units are selected by the pattern-representing lateral connections (which may be called "horizontal nets"), thus factoring in memory patterns, the winner unit inside feature columns are selected by another kind of net structure, "vertical nets", which are formed by connections running between value units in different sub-modalities. A vertical net ties together feature value units that are consistent with each other, consistent in the sense of signals arriving at a unit over alternate pathways within the net as well as sensory signals agree with each other. Like the net structures representing memory for local feature distribution, consistency nets are established by a combination of learning from sensory input and self-interaction.

# INTRINSIC COORDINATE DOMAIN

So far, we have spoken of structure in primary visual cortex, which is dominated by retinal coordinates, that is, image location changes with eye movements. All local texture representation must therefore be repeated for all positions. (This is possible only for a limited number of local texture patches, comparable to the sizes of codebooks in image-compression schemes). In order to store and represent larger chunks of visual structure, such as for recurring patterns like familiar objects or abstract whole-scene lay-outs, there is another domain, see Figure 2, presumably infero-temporal cortex, in which neurons, units and columns refer to pattern-fixed, intrinsic coordinates. (For the structure of fibre projections between the retinal-coordinate and the intrinsic-coordinate domains see below.) The intrinsic domain can be much more parsimonious then the retinal one in not needing to repeat net structures over the whole visual field, so that it can afford to spend more redundancy in each intrinsic location to be able to store a very large number of pattern-spanning nets.

Also the intrinsic domain contains sub-structures for the representation of sub-modalities, and again there are nets for the representation of mutual constraints between the sub-modalities. Thus, the two domains are qualitatively the same but quantitatively very different.

# DYNAMIC MAPPINGS

The two domains with retinal and intrinsic coordinates are connected by dynamic point-to-point and feature-to-feature fibre projections that can be switched as quickly as retinal images move, so that correspondence between homeomorphic structures is maintained. This switching is achieved with the help of "control units" (Anderson and VanEssen, 1987). These can be realized as neurons whose outgoing synapses are co-localized with the synapses of the projection fibres they control at dendritic patches of the target neurons. If those patches have threshold properties, the projection fibres can transmit signals only if also the controlling fibre is active. The hypothesis that dendritic patches with non-linear response properties are act as decision units has been proposed long ago, see for instance (Polsky et al. 2004).

proportional to the similarity of the signal pattern in the controlled projection fibres on the one hand and the signal pattern in the target neurons on the other. (Processes of control neurons would thus transmit *and* receive signals and should correspondingly be called neurites.)

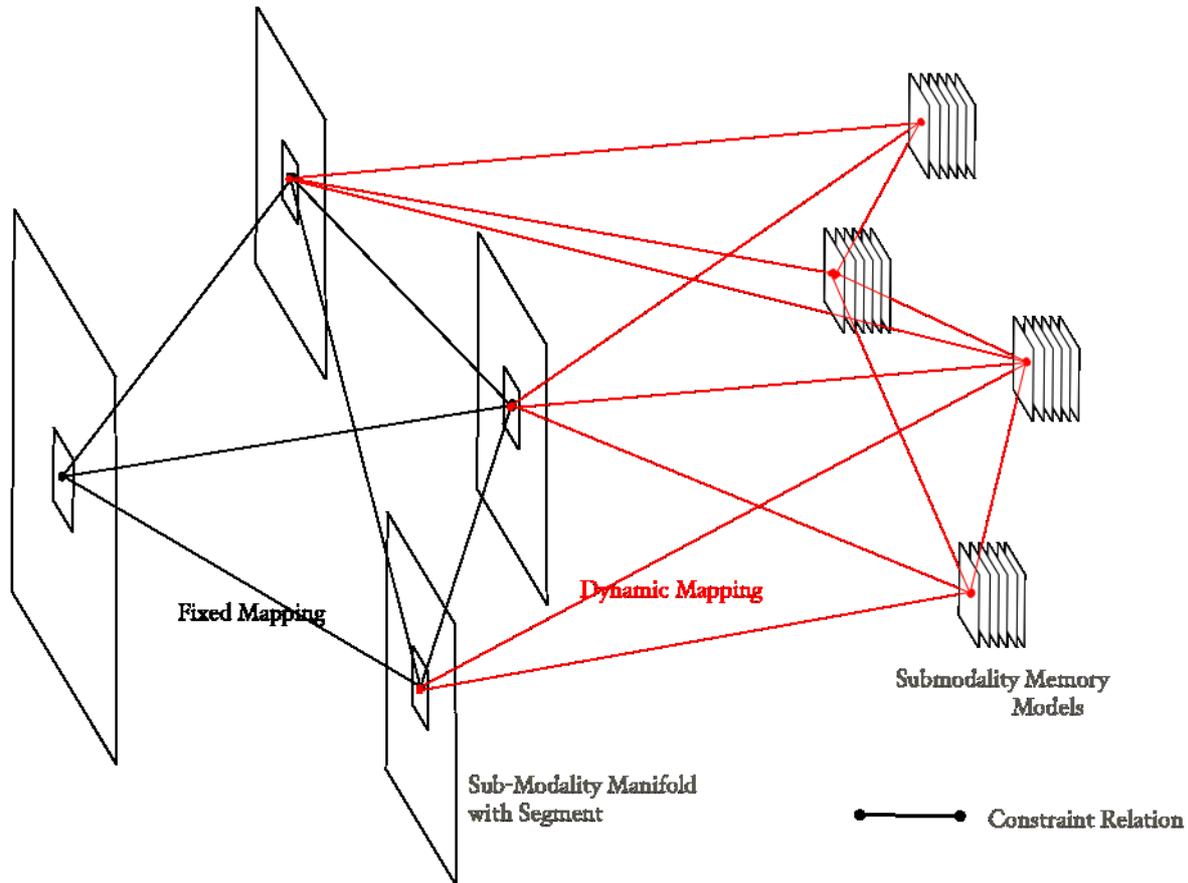

Figure 2: Overview of the Architecture. Each plane corresponds to one sub-modality, on the left side in retinal coordinates (primary visual cortex), on the right side in pattern-intrinsic coordinates (infero-temporal cortex). A segment in the retinal-coordinate domain is projected by dynamical mappings to the intrinsic-coordinate domain. Constraint interactions that help to single out mutually latent variable values run between corresponding points (which refer to the same point on a surface within the visual scene).

A control neuron may, like any other neuron, receive synaptic inputs (e.g., by re-afferent signals that can in this way switch projection fibres such as to compensate an intended eye movement), but they also may get excited through their control fibres. We assume that these carry signals that are

Different control units stand for different transformation parameters (relative position, size or orientation of connected sets of neurons in the two domains) and may be responsible for connecting a local patch in one domain to a local patch in the other. The set of control units for different transformation parameters for a given patch in the target domain form a column with WTA dynamic and represents transformation parameters as latent variable. In order to cover deformation, transformation parameters may change slowly from point-to-point in the target domain, and an entire coherent mapping is represented by a net of laterally connected control units (again, units contain a number of redundant neurons to give leeway for many nets to be stored side-by-side without mutual interference).

## SEGMENTATION

Vision is organized as a sequence of attention flashes. During each such flash, analysis of sensory input is restricted to a segment – a coherent chunk of structure – e.g., to the region in retinal space that is occupied by the image of an object. Like perception in general, segmentation is a hen-and-egg problem, segmentation needing recognition, recognition needing segmentation. Certain patterns indicative of a coherent structure are already available in primary cortex, such as the presence of coherent fields of motion, depth or colour, or familiar contour shapes. Others, however, need reference to patterns stored in the intrinsic coordinate domain. For this to happen, two types of latent variables have to be made to converge first, the transformation parameters identifying the segment's location and size in the retinal-coordinate domain, and the identity of structure of a fitting model in memory. We have modelled this process for the purpose of object recognition, which we tested successfully on a benchmark, observing rather fast convergence (reference suppressed for anonymity) . In general, the intrinsic representation of the segment cannot be found in memory but needs to be assembled from partial patterns (just as the extended texture in primary cortex is assembled from local texture patches). Conceiving of objects as composites of known elementary shapes is a well-established concept (Biederman, 1987). This process of assembly takes place in a coordinated fashion in the different sub-modality modules. The de-composition and re-composition of sensory patterns is the basis for a very parsimonious system of representing a large combinatorial universe of surfaces of different shape, colouring texture under a range of illumination conditions and in different states of motion.

## RECOGNITION

The actual recognition process of a pattern in the retinal coordinate domain against a pattern in the intrinsic domain may be seen as a process of finding a homeomorphic projection, or graph matching, performed by many control neurons simultaneously checking for patterns similarity while competing with alternate control neurons and cooperating with compatible ones (compatible in the sense of forming together a net structure). Recognition by graph matching has a long tradition, see for instance (Kree and Zippelius, 1988) or (Lades et al., 1993). A related approach is (Arathorn, 2002), who has pointed out the value of the information inherent in the shape of the mapping, which is produced as a by-product.

## PREDICTION

Once this process has converged for a moving pattern and its motion parameters have also been determined, the system can set the fibre projection system in motion to track the object and send short-term predictions of sensory input down from the model in the intrinsic domain to the primary cortex. Successful prediction of sensory input on the basis of a constructed dynamic model is the ultimate basis for our confidence in perceptual interpretations of the environment, and is very important for the adjustment of constraint interactions.

## ONGOING WORK AND NEXT STEPS

We are at present working on a simple version of the architecture, implementing the modalities grey-level (the input signal), surface reflectance, illumination, depth, surface orientation and shading, all realized in image coordinates. We are manually creating constraint interactions between them and a small number of lateral connectivity nets. The goal is to model the perceptual collapse on simple sample images. To suppress the tendency of the system to break up spontaneously into local domains (generating spurious latent-variable discontinuities) we are working with coarse-to-fine strategies. As we are embedded in a lab that is engaged in an effort to build a computer vision system by methods of systems engineering, we plan to adapt more and more known vision algorithms into the architecture.

## CONCLUSION

All we are proposing is to re-interpret neural tissue and dynamics such as to see them as the natural basis for the structures and processes that are required for vision. The essential point is the assumption that neural tissue is an overlay of well-structured "nets", which are characterized by sparsity in terms of connection per neuron and consistency of different pathways between pairs of neurons. Particular supporting assumptions are exploitation of cell-number redundancy and winner-take-all dynamics to disentangle different nets and to represent latent variables, and co-localization and non-linear interaction of synapses on dendritic patches. Nets are activated on the perceptual timescale and are generated by self-interaction and Hebbian plasticity on the learning timescale. Nets act as data structure for the representation of memory patterns, as networks of constraints between latent variables in different sub-modalities and as projection fibre mappings between retinal and intrinsic coordinate systems. The coherent architecture that is shaped by these assumptions promises to decisively expand the functional repertoire of neural models. This architecture may even help to unify the as yet very heterogeneous array of algorithms and data structures that has arisen in computer vision, an urgent precondition for progress in that field.

There is an important type of experimental prediction flowing from our proposal concerning the detailed wiring diagram of cortical tissue. Analogous to the network of molecular interactions, which is dominated by "motifs" (Shen-Orr et al., 2002), connectivity should be dominated by closed loops or "diamond motifs", short alternate pathways starting in one neuron and ending in another, or even on the same dendritic patch of another neuron. This may turn out to be a very important type of results in the upcoming era of connectomics.